\documentclass[prd,preprint,showpacs]{revtex4}
\usepackage{amsmath}
\usepackage{latexsym}
\usepackage{amsfonts}
\usepackage[dvips]{graphicx}
\newcommand{\be}{\begin{equation}}
\newcommand{\ee}{\end{equation}}
\begin{document}
\title{The Holographic dark energy reexamined}
\author{Yungui Gong} \email{gongyg@cqupt.edu.cn}
\affiliation{College of Electronic Engineering, Chongqing
University of Posts and Telecommunications, Chongqing 400065,
China}
\author{Bin Wang}
\email{wangb@fudan.edu.cn} \affiliation{ Department of Physics,
Fudan University, Shanghai 200433, China}
\author{Yuan-Zhong Zhang}
\affiliation{ CCAST (World Laboratory), P.O. Box 8730, Beijing 100080 \\
Institute of Theoretical Physics, Chinese Academy of Sciences,
P.O. Box 2735, Beijing 100080, China}
\begin{abstract}
We have reexamined the holographic dark energy model by
considering the spatial curvature. We have refined the model
parameter and observed that the holographic dark energy model does
not behave as phantom model. Comparing the holographic dark energy model
to the supernova observation alone, we found that the closed
universe is favored. Combining with the Wilkinson Microwave
Anisotropy Probe (WMAP) data, we obtained the reasonable value of
the spatial curvature of our universe.
\end{abstract}
\preprint{hep-th/0412218}
\maketitle

\parindent=4ex

\section{Introduction}
The total entropy of matter inside a black hole cannot be greater
than the Bekenstein-Hawking entropy, which is one quarter of the
area of the event horizon of the black hole measured in Planck
unit. In view of the example of black hole entropy, Bekenstein
proposed a universal entropy bound $S\le 2\pi ER$ for a weak
self-gravitating physical system with total energy $E$ and size
$R$ in 1981 \cite{bekenstein}. Later 't Hooft and Susskind
proposed an influential holographic principle, relating the
maximum number of degrees of freedom in a volume to its boundary
surface area\cite{holography}. The extension of the holographic
principle to the cosmological setting was first addressed by
Fischler and Susskind (FS) \cite{fischler}. Subsequently, various
 modifications of the FS version of the holographic
principle was proposed \cite{bousso}. The idea of the
holographic principle is viewed as a real conceptual change in our
thinking about gravity \cite{witten}. It has appeared many
examples of applying the holographic principle to study cosmology,
such as understanding the possible value of the cosmological
constant \cite{cc}\cite{cohen}, selecting physically acceptable
model in inhomogeneous cosmology \cite{inhom} and discussing upper
limits on the number of e-foldings in inflation \cite{efolding}.
It is of great interest to generalize the application of
holography to a much broader class of situations, especially to
cosmology.

The type Ia supernova (SN Ia) observations suggest that the
Universe is dominated by dark energy with negative pressure which
provides the dynamical mechanism of the accelerating expansion of
the Universe \cite{sp99,agr98}. The simplest candidate of
dark energy is the cosmological constant. However the unusual
small value of the cosmological constant is a big challenge to
theoretical physicists. Whether holography can shed us
some light in understanding the profound puzzle posed by the dark
energy is a question we want to ask. Motivated by the assumption that for any
state in the Hilbert space with energy $E$, the corresponding
Schwarzschild radius $R_s\sim E$ is less than the infrared (IR)
cutoff $L$ \cite{cohen},
 a relationship between the
ultraviolet (UV) cutoff and the infrared cutoff is derived, i.e.,
$8\pi GL^3\rho_D/3\sim L$ \cite{cohen}. We can express the holographic dark
energy density as
\begin{equation}
\label{hldark} \rho_D= \frac{3c^2\,d^2}{8\pi G L^2},
\end{equation}
where $c$ is the speed of light and $d$ is a constant of the order
of unity. This UV-IR relationship was also obtained by Padmanabhan by arguing that
the cosmological constant is due to the vacuum fluctuation of energy density.
Hsu found that the holographic dark energy model based
on the Hubble scale as the IR cutoff won't give an accelerating
universe \cite{hldark2}. In \cite{hldark3}, Li showed that
choosing the particle horizon as the IR cutoff,  an accelerating
universe will not be produced either. However, by relating the IR
cutoff to an event horizon, it was found that the holographic dark
energy model can accommodate the accelerating universe
\cite{hldark3,huang2}. The model in the flat universe was found in
consistent with current observations \cite{huang1}. Here we would like to
point out that the form
$\rho_D\sim H^2$ also works for dark energy model building. For example,
the model $\rho_D=\rho_\Lambda+3c^2d^2 H^2/(8\pi G)$ with $\rho_\Lambda$ a constant derived from the
re-normalization group models of the cosmological constant can explain the
accelerating expansion of the Universe \cite{hubble}. Ito also discovered a viable
holographic dark energy model by using the Hubble scale as the IR
cutoff with the use of non-minimal coupling to scalar field
\cite{ito}. More recently, a dark energy model $\rho_D\sim H^2$ with an interaction
between the dark energy and dark matter was proposed to explain the accelerating expansion \cite{pavon}. The
holographic dark energy model in the framework of Brans-Dicke
theory was discussed in \cite{gong04}. Some speculations about the
deep reasons of the holographic dark energy were considered by
several authors \cite{origin}. The holographic principle was also
used to constrain dark energy models in \cite{wang}. In this
paper, we reexamine the holographic dark energy model proposed
in \cite{hldark3}. We give constraints on this model from
both the theoretical argument and the observational data. Including
the spatial curvature, we will find that the closed universe is
marginally favored. This result agrees to the Cosmic
Microwave Background (CMB) Anisotropy experiments \cite{cmb,wmap1,smalll} and
recent supernova investigations \cite{supernova}.

\section{Holographic dark energy model with curvature}
We start from the homogeneous and isotropic Friedmann-Robertson-Walker (FRW)
space-time metric
\begin{equation}
\label{rwcosm} ds^2=-c^2dt^2+a^2(t)\left[{dr^2\over
1-k\,r^2}+r^2\,d\Omega\right].
\end{equation}
If a light is emitted from a point $r_1$ at time $t_1$, it will arrive at the origin at time $t_0$.
The light follows the null geodesics, so we have \be \label{line}
\int_{t_1}^{t_0}{c\,dt\over a(t)}=\int_0^{r_1}{dr\over
\sqrt{1-kr^2}}\equiv f(r_1), \ee where
\begin{eqnarray*}
f(r_1)&=&\frac{1}{\sqrt{|k|}}{\rm
sinn}^{-1}(\sqrt{|k|}\,r_1)\nonumber\\
&=&\left\{\begin{array}{ll}
\sin^{-1}(\sqrt{|k|}\,r_1)/\sqrt{|k|},\ \ \ \ \ \ &k=1,\\
r_1,&k=0,\\
\sinh^{-1}(\sqrt{|k|}\,r_1)/\sqrt{|k|},&k=-1.
\end{array}\right.
\end{eqnarray*}
With both an ordinary pressureless dust matter and the holographic
dark energy as sources, the Friedmann equations are
\begin{gather}
\label{cos1} H^2+{kc^2\over a^2}={8\pi
G\over 3}(\rho_m+\rho_r+\rho_D),\\
\label{cos2} \dot{\rho_D}+3H(\rho_D+p_D)=0,
\end{gather}
where the Hubble parameter $H=\dot{a}/a$,  the matter density
$\rho_m=\rho_{m0}(1/a)^3$, the radiation density
$\rho_r=\rho_{r0}(1/a)^4$, the dot means derivative with respect
to time and the subscript 0 means the value of the variable at
present time and $a_0=1$ is set.

Now as done in \cite{hldark3} we  choose the event horizon as the IR cutoff, where
\begin{gather} \label{reh} R_{eh}(t)=a(t)\int^\infty_t
\frac{cdt}{a(t)}=a(t)\int^\infty_{a(t)}
\frac{cd\tilde{a}}{\tilde{a}^2H}=\int_0^r{d\tilde{r}\over
\sqrt{1-k\tilde{r}^2}},\\
\label{leh} L=a(t)r=\frac{a(t) {\rm
sinn}[\sqrt{|k|}\,R_{eh}(t)/a(t)]}{\sqrt{|k|}}.\end{gather}
Apparently, we recover $L=R_{eh}$ for a spatially flat universe.

Let us rewrite Eq. (\ref{cos1}) as \be \label{frweq}
\Omega_m+\Omega_r+\Omega_D=1+\Omega_k, \ee where
$\Omega_m=\rho_m/\rho_{cr}=\Omega_{m0}H^2_0/(H^2a^3)$,
$\Omega_r=\rho_r/\rho_{cr}=\Omega_{r0}H^2_0/(H^2a^4)$,
$\Omega_D=d^2 c^2/(L^2 H^2)$ and
$\Omega_k=kc^2/(a^2H^2)=\Omega_{k0}H^2_0/(a^2H^2)$. Since
$$\frac{\Omega_k}{\Omega_m}=a\frac{\Omega_{k0}}{\Omega_{m0}}=a\gamma,$$
where $\gamma=\Omega_{k0}/\Omega_{m0}$, and
$$\frac{\Omega_r}{\Omega_m}=\frac{\Omega_{r0}}{a\Omega_{m0}}=\frac{\beta}{a},$$
where $\beta=\Omega_{r0}/\Omega_{m0}=1/(1+z_{eq})$ and the matter
radiation equality redshift $z_{eq}=3233$ \cite{wmap}, we have \be
\label{wmeq}
\Omega_m=\frac{\Omega_{m0}H^2_0}{H^2a^3}=\frac{a(1-\Omega_D)}{\beta+a-a^2\gamma}.
\ee From the above equation, we get \be \label{wmeq1}
\frac{1}{aH}=\frac{a}{H_0}\sqrt{\frac{1-\Omega_D}{\Omega_{m0}(\beta+a-a^2\gamma)}}.
\ee
Combining Eqs. (\ref{leh}) and (\ref{wmeq1}) and using the
definition of $\Omega_D$, we obtain
\begin{eqnarray} \label{wleq} \sqrt{|k|}\frac{R_{eh}}{a}&=&{\rm
sinn}^{-1}\left[d\sqrt{|\gamma|}\sqrt{\frac{a^2(1-\Omega_D)}{\Omega_D(\beta+a-a^2\gamma)}}\,\right]\nonumber\\
&=&{\rm sinn}^{-1}(d\,\sqrt{|\Omega_k|/\Omega_D}).
\end{eqnarray}
If $\Omega_k>0$, then we require $d\le
\sqrt{\Omega_D/\Omega_k}$.

By using Eqs. (\ref{hldark}), (\ref{cos2})-(\ref{leh}) and
(\ref{wleq}), we get the dark energy equation of state
\begin{eqnarray} \label{ehol}
w_D&=&-\frac{1}{3}\frac{d\ln\rho_D}{d\ln
a}-1\nonumber\\
&=&-\frac{1}{3}\left[1+\frac{2}{d}\sqrt{\Omega_D}{\rm
cosn}(\sqrt{|k|}\,R_{eh}/a)\right]\nonumber\\
&=&-\frac{1}{3}\left[1+\frac{2}{d}\sqrt{\Omega_D-d^2\Omega_k}\right],
\end{eqnarray}
where
\begin{equation*}
\frac{1}{\sqrt{|k|}}{\rm cosn}(\sqrt{|k|}x)
=\left\{\begin{array}{ll}
\cos(x),\ \ \ \ \ \ &k=1,\\
1,&k=0,\\
\cosh(x),&k=-1.
\end{array}\right.
\end{equation*}
It is obvious that $w_D\le -1/3$, so we can have an
accelerating universe.

Taking derivative with respect to $a$ on both sides of Eq.
(\ref{wleq}) and use the redshift $z=1/a-1$ as the variable, we
get the following differential equation by using Eqs. (\ref{reh})
and (\ref{wmeq1})
\begin{eqnarray} \label{wldeq}
\frac{d\Omega_D}{dz}&=&-\frac{2\Omega_D^{3/2}(1-\Omega_D)}{d(1+z)}
\sqrt{1-\frac{d^2\gamma(1-\Omega_D)}{\Omega_D[\beta(1+z)^2+1+z-\gamma]}}\nonumber\\
&&-\frac{\Omega_D(1-\Omega_D)[1+2\beta(1+z)]}{\beta(1+z)^2+1+z-\gamma}.
\end{eqnarray}
With this expression, we can understand the evolution behavior of the dark energy.

Now let us find the constraints on the parameter $d$ in the holographic dark energy model.
The entropy of the whole system is described
by $S=\pi M_p^2 L^2$. To satisfy the second law of thermodynamics,
we require that
\begin{eqnarray} \label{lt} \dot{L}&=&LH-c\,{\rm
cosn}[\sqrt{|k|}\,R_{eh}(t)/a(t)]\nonumber\\
&=&c\left(\frac{d}{\sqrt{\Omega_D}}-\sqrt{1-\frac{d^2\gamma(1-\Omega_D)}
{\Omega_D[\beta(1+z)^2+1+z-\gamma]}}\right) \nonumber\\&\ge&
0,
\end{eqnarray}
 Thus
 \begin{equation}
 \label{dlbnd}
 d^2\ge \frac{\Omega_D[\beta(1+z)^2+1+z-\gamma]}{\beta(1+z)^2+1+z-\gamma\Omega_D}=\frac{\Omega_D}{1+\Omega_k}.
 \end{equation}
For the spatially flat universe, we recover $d^2\ge
\Omega_D$. When the dark energy dominates, $d^2\ge 1$, which is the lower bound of $d$ proposed in \cite{huang2}.

In addition to the lower bound on $d$, employing the argument that
the total energy in a region of size $L$ should not exceed the
mass of a black hole of the same size, we have the upper bound
$d\le 1$. Alternatively $d\le 1$ can be argued by using the
condition $R_s\le L$. For a dark energy dominated universe, we
have
\begin{equation}
\label{rs1}
R_s=\frac{2GM}{c^2}=2G\rho_D\left(\frac{4\pi}{3c^2}L^3\right)\le
L,
\end{equation}
so
\begin{equation}
\label{hlwl} \rho_D\le \frac{3c^2}{8\pi G L^2}.
\end{equation}
Comparing Eqs. (\ref{hldark}) and (\ref{hlwl}), we get $d\le 1$.
Thus we find that $d$ must lie in the range
\begin{equation} \label{drange}
\sqrt{\frac{\Omega_D}{1+\Omega_k}}\le d\le 1.
\end{equation}
 As the dark energy gradually dominates the universe,
$\Omega_D\rightarrow 1$, the allowed range of $d$ will
become smaller. It is also interesting to note that the Bekenstein
entropy bound
\begin{equation}
\label{sbound} S\le \frac{2\pi E L}{c}= \frac{8\pi^2 c
\rho_D L^4}{3}\le\frac{\pi c^3 L^2}{G}=S_{\rm BH}.
\end{equation}
Therefore, the maximum entropy is the Bekenstein-Hawking entropy
$S_{\rm BH}$.

Applying the constraint Eq. (\ref{drange}) to Eq. (\ref{ehol}), we
find that $w_D\ge -1$. Therefore, the holographic dark
energy has no phantom-like behavior.

\section{Phenomenological Consequences}
Now we use the 157 gold sample SN Ia data compiled in
\cite{riess04} to fit the model. The parameters $d$, $\Omega_{m0}$
and $\Omega_{k0}$ in the model are determined by minimizing
\be\label{chi2} \chi^2=\sum_i{[\mu_{\rm obs}(z_i)-\mu(z_i)]^2\over
\sigma^2_i},\ee where the extinction-corrected distance modulus
$\mu(z)=5\log_{10}(d_{\rm L}(z)/{\rm Mpc})+25$, the luminosity
distance is
\begin{eqnarray}
d_{\rm L}&=&(1+z)r(z)\nonumber\\
&=&\frac{c(1+z)}{H_0\sqrt{|\Omega_{k0}|}}\,{\rm sinn}
(\sqrt{|k|}[(1+z)R_{eh}(z)-R_{eh}(0)])\nonumber\\
&=&\frac{c(1+z)}{H_0\sqrt{|\Omega_{k0}|}}\,{\rm sinn}\left[-{\rm
sinn}^{-1}\left(\sqrt{\frac{d^2|\Omega_{k0}|}{\Omega_{D0}}}\right)\right.\nonumber\\
&&\left.+{\rm
sinn}^{-1}\left(\sqrt{\frac{d^2|\gamma|(1-\Omega_D)}
{\Omega_D[\beta(1+z)^2+1+z-\gamma]}}\right)\right],
\end{eqnarray}
$\sigma_i$ is the total uncertainty in the observation. The
nuisance parameter $H_0$ is marginalized over with a flat prior
assumption. Since $H_0$ appears linearly as the form of
$5\log_{10}H_0$ in $\chi^2$, the marginalization by integrating
${\mathcal L}=\exp(-\chi^2/2)$ over all possible values of $H_0$ is
equivalent to finding the value of $H_0$ which minimizes $\chi^2$
if we also include the suitable integration constant. Therefore we
marginalize $H_0$ by minimizing $\chi^{\prime
2}=\chi^2(y)-2\ln(10)\,y/5-2\ln[\ln(10)\sqrt{(2\pi/\sum_i
1/\sigma^2_i)}/5]$ over $y$, where $y=5\log_{10}H_0$. We also
assume a prior $\Omega_{m0}=0.3\pm 0.1$. The parameter space for
$\Omega_{m0}$ is $[0,\ 1]$, the parameter space for $\Omega_{k0}$
is $[-1,\ 1]$ and the parameter space for $d$ is coming from the
constraint Eq. (\ref{drange}). The best fit parameters are
$\Omega_{m0}=0.35^{+0.11}_{-0.10}$,
$\Omega_{k0}=0.35^{+0.17}_{-0.38}$ and $d=1.0_{-0.17}$ with
$\chi^2=173.35$. Note that $d$ has reached the upper bound 1, so
there is no positive error for $d$. The error is referred to $1\sigma$ error throughout this paper.
For the flat universe, the best fit parameters
are $\Omega_{m0}=0.30^{+0.04}_{-0.08}$ and
$d=0.84^{+0.16}_{-0.03}$ with $\chi^2=176.33$. For comparison, the
best fit to the flat $\Lambda$CDM model gives $\chi^2=176.51$. Therefore
using the holographic dark energy model from the supernova data
fitting, the closed universe is marginally favored compared to the
flat case.

\begin{figure}[htb]
\begin{center}
\includegraphics[width=0.5\textwidth]{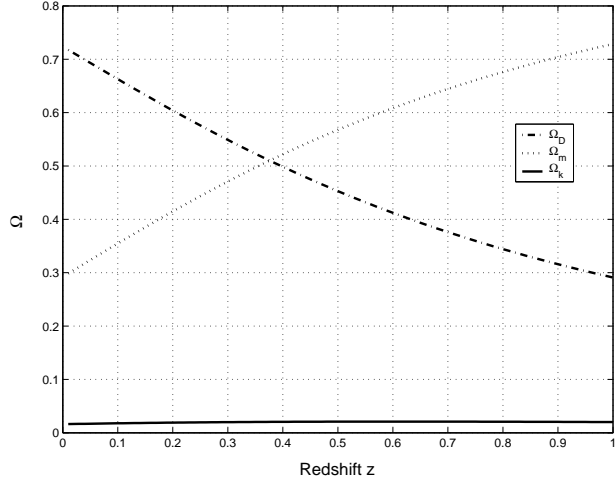}
\end{center}
 \vspace*{-0.3in} \caption{The evolution of
$\Omega_D$, $\Omega_m$ and $\Omega_k$ by using the best fit
parameters $\Omega_{m0}=0.29$, $\Omega_{k0}=0.02$ and $d=0.84$.}
\label{holoomga}
\end{figure}

\begin{figure}[htb]
\begin{center}
\includegraphics[width=0.5\textwidth]{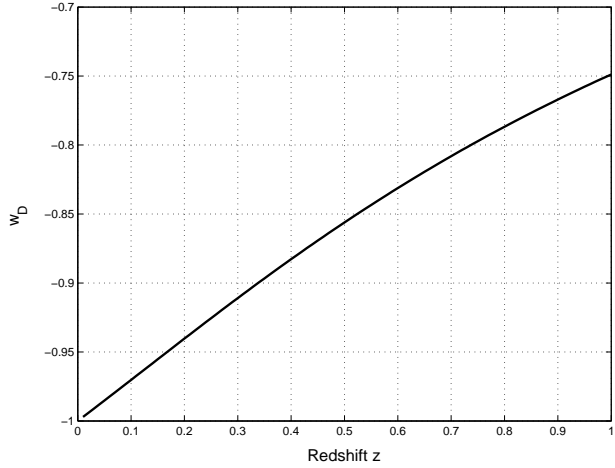}
\end{center}
 \vspace*{-0.3in} \caption{The evolution of
$w_D$ by using the best fit parameters $\Omega_{m0}=0.29$,
$\Omega_{k0}=0.02$ and $d=0.84$.} \label{holowl}
\end{figure}

To further
constrain the model, we combine the SN Ia data with the WMAP data. The main effect of changing the values
of $\Omega_{m0}$ and $\Omega_{k0}$ on the CMB anisotropy can be found from the shift
parameter $\mathcal{R}$ with which the $l$-space positions of the acoustic peaks
in the angular power spectrum shift \cite{shift},
\begin{eqnarray}
\label{shift} \mathcal{R}&=&\sqrt{\Omega_{m0}}H_0r(z_{ls})/c
\nonumber\\
&=&\frac{1}{\sqrt{|\gamma|}}\,{\rm sinn}\left[-{\rm
sinn}^{-1}\left(\sqrt{\frac{d^2|\Omega_{k0}|}{\Omega_{D0}}}\right)\right.\nonumber\\
&&\left.+{\rm
sinn}^{-1}\left(\sqrt{\frac{d^2|\gamma|(1-\Omega_D)}
{\Omega_D[\beta(1+z_{ls})^2+1+z_{ls}-\gamma]}}\right)\right]\nonumber\\
&=&1.710\pm 0.137,
\end{eqnarray}
where $z_{ls}=1089\pm 1$ \cite{wmap}. Therefore we use the above shift parameter along with the SN Ia data
to fit the model.  The best
fit parameters are $\Omega_{m0}=0.29^{+0.06}_{-0.08}$,
$\Omega_{k0}=0.02\pm 0.10$ and $d=0.84^{+0.16}_{-0.03}$ with
$\chi^2=176.12$. It is interesting to note that this best fitting result
presents us the same curvature of the universe as that from the WMAP observation. This result suggests
that the WMAP data prefers an almost spatially flat universe while the SN Ia data gives a closed universe.
By using the best fit parameters, we plot the
evolutions of $\Omega_D$, $\Omega_m$ and $\Omega_k$ in Fig. \ref{holoomga}.
From Fig. \ref{holoomga}, we see that $\Omega_D\rightarrow
1$, $\Omega_m\rightarrow 0$ and $\Omega_k\rightarrow
-1+\Omega_D=0$. Combining Eqs. (\ref{wldeq})
and (\ref{ehol}), we get the evolution of $w_D$. The result
is plotted in Fig. \ref{holowl}.
From Fig. \ref{holowl}, we see that as expected the holographic
dark energy does not have phantom like behavior.

\begin{figure}[htb]
\begin{center}
\includegraphics[width=0.5\textwidth]{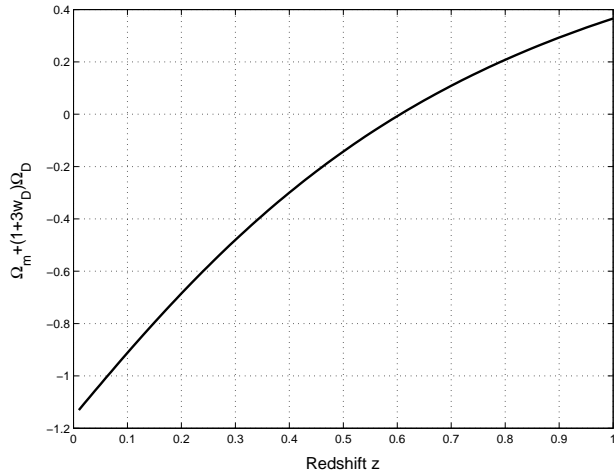}
\end{center}
 \vspace*{-0.3in} \caption{The evolution of
$-2\ddot{a}/(aH^2)$ by using the best fit
parameters $\Omega_{m0}=0.29$, $\Omega_{k0}=0.02$ and $d=0.84$.}
\label{holoacc}
\end{figure}

\begin{figure}[htb]
\begin{center}
\includegraphics[width=0.5\textwidth]{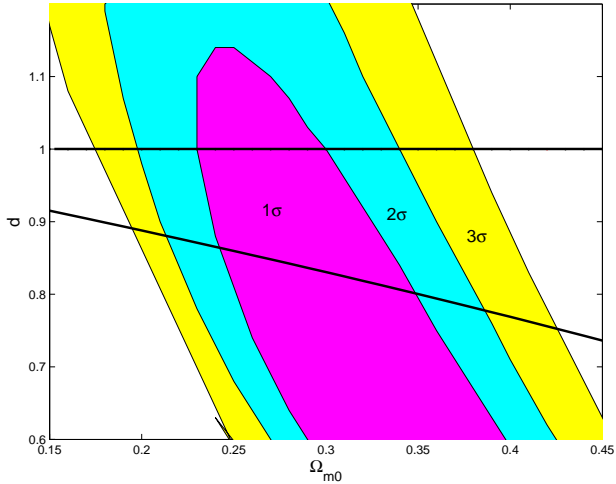}
\end{center}
 \vspace*{-0.3in} \caption{The $1\sigma$, $2\sigma$ and $3\sigma$ contour plots for
 $\Omega_{m0}$ and $d$ by using $\Omega_{k0}=0.02$. The contours are those regions intersecting with the two black lines
 due to the constraint Eq. (\ref{drange}).}
\label{contfig}
\end{figure}

Using Eqs. (\ref{cos1}) and (\ref{cos2}), we get the acceleration equation
\begin{eqnarray}
\frac{\ddot{a}}{a}&=&-\frac{4\pi G}{3}(\rho_m+\rho_D+3p_D)\nonumber\\
&=&-\frac{H^2}{2}[\Omega_m+(1+3w_D)\Omega_D].
\end{eqnarray}
It is clear that the sign of $\Omega_m+(1+3w_D)\Omega_D$ determines the sign of $\ddot{a}$. Combining the
behaviors of $\Omega_D$, $\Omega_m$ and $w_D$, we plot
the evolution of $\Omega_m+(1+3w_D)\Omega_D=-2\ddot{a}/(aH^2)$ which
shows the behavior of acceleration in Fig. \ref{holoacc}.
From Fig. \ref{holoacc}, we see that the universe experienced the
transition from deceleration to acceleration around $z_t=0.6$. By fixing $\Omega_{k0}$ at its best fit
value $\Omega_{k0}=0.02$, we give the contour plot for $\Omega_{m0}$ and $d$ in Fig. \ref{contfig}. For
the spatially flat holographic model, the best fit parameters are
$\Omega_{m0}=0.28\pm 0.05$ and $d=0.85^{+0.15}_{-0.03}$ with
$\chi^2=176.18$. Again, for comparison, the best fit parameter of
the flat $\Lambda$CDM model is $\Omega_{m0}=0.31^{+0.04}_{-0.03}$ with
$\chi^2=176.61$. Thus combining with the WMAP data, the closed universe still cannot be ruled out.

\section{Conclusions}
In conclusion, we have reexamined the holographic dark energy
model and given a constraint on its parameter. By comparing
to observations, we found that the holographic model is an
effective model in describing dark energy. A spatially closed
universe is favored by using the SN Ia data alone. Combining with
the WMAP data, the best fitting result gives us a reasonable value
of the curvature of our universe and the closed universe cannot be
ruled out. Statistically the closed universe plays the same role
as the flat universe in comparing with observations. By
investigating the evolution of the dark energy, we observed that
the transition of our universe from the deceleration to the
acceleration happens at $z_t=0.6$. In Ref. \cite{huang1}, one of
us discussed the spatially flat holographic dark energy model and
found that $\Omega_{m0}=0.46$ and $d=0.20$, the model behaved like
phantom. In this paper, we used the arguments of the second law of
thermodynamics and the holographic principle to get the lower and
upper bounds on the parameter $d$. Due to the constraint Eq.
(\ref{drange}), the holographic model discussed in this paper has
no phantom-like behavior. Furthermore, we get a lower value of
$\Omega_{m0}$ which is more consistent with other observations on
the value of the non-relativistic matter energy density.

Comparing with Ref. \cite{huang1}, we have included the curvature
of the universe in our discussion. The SN Ia data alone favors the
closed universe with a bit bigger $\Omega_k$, while combining with
the WMAP data, $\Omega_k$ decreases to a value around $0.02$. This
discussion is not trivial. Although our result is consistent with
the viewpoint that our universe is approximately flat, the small
curvature of the universe is still interesting since it may
contribute to the small $l$ suppress of the CMB spectrum
\cite{smalll}.

\begin{acknowledgments}
This work was initiated in the workshop held in the SIAS-USTC. Y.
Gong would acknowledge helpful discussion with R.G. Cai and his
work was supported by CQUPT under Grant No. A2004-05, CSTC under
grant No. 2004BB8601, SRF for ROCS, State Education Ministry
and NNSFC under grant No. 10447008. B. Wang's
work was partially supported by NNSFC, Ministry of Education of
China and Ministry of Science and Technology of China under grant
NKBRSFG19990754. Y.Z. Zhang's work was in part supported by NNSFC
under Grant No. 90403032 and also by National Basic Research
Program of China under Grant No. 2003CB716300.

\end{acknowledgments}

\end{document}